\documentclass{article}
\usepackage{graphicx}

\usepackage{amssymb,amsthm,amsmath}
\usepackage{xcolor,paralist,hyperref,titlesec,fancyhdr,etoolbox}

\titlespacing*{\section}{0pt}{0ex}{0ex}

\hypersetup{ colorlinks=true, linkcolor=black, filecolor=black, urlcolor=black }

\usepackage{lipsum}

\begin{document}

\begin{flushleft}
{\Large
\textbf\newline{Active Encoding of Flexural Wave with Non-Diffractive Talbot Effect}
}
\newline
\\
Zhiqiang Li \textsuperscript{1},
Kaiming Liu \textsuperscript{1},
Chunlin Li \textsuperscript{2},
Yongquan Liu \textsuperscript{2},
Ting Li \textsuperscript{1},
Zhaoyong Sun \textsuperscript{1,$^a$},
Liuxian Zhao \textsuperscript{3,$^b$},
Jun Yang \textsuperscript{4,$^c$}
\\
\bigskip
1 Beijing Institute of Graphic Communication,1 Xinghua Avenue (Band 2), Beijing, 102600,China
\\
2 State Key Laboratory for Strength and Vibration of Mechanical Structures, Department of Engineering
Mechanics, School of Aerospace Engineering, Xi’an Jiaotong University, Xi’an 710049, China
\\
{3} Institute of Sound and Vibration Research, Hefei University of Technology, 193 Tunxi Road, Hefei 230009, China
\\
{4} Key Laboratory of Noise and Vibration Research, Institute of Acoustics, Chinese Academy of Sciences, 21 North 4th Ring Road, Beijing 100190, China
\\
\bigskip
$^a$ sunzhaoyong@bigc.edu.cn 

$^b$ lxzhao@hfut.edu.cn 

$^c$ jyang@mail.ioa.ac.cn

\end{flushleft}

\begin{abstract}

This study employs the theory of conformal transformation to devise a Mikaelian lens for flexural waves manipulation. 
We investigate the propagation patterns of flexural waves in the lens under scenarios of plane wave and point source incidence. 
Additionally, the study explores the Talbot effect generated by interference patterns of multiple sources. 
Within the Mikaelian lens, the Talbot effect displays non-diffractive characteristics, facilitating propagation over considerable distances. 
Leveraging the non-diffractive attributes of the Talbot effect in the Mikaelian lens, the paper discusses the feasibility of encoding flexural waves based on active interference sources. 
Simulation and experimental validation attest to the lens's effective active encoding. 
This research introduces novel perspectives on flexural wave encoding, showcasing potential applications in flexural wave communication, detection, and related fields.

\end{abstract}

\section{Introduction}
Metamaterials have experienced rapid development in recent decades due to their advantages, such as subwavelength dimensions and flexible control of physical parameters\cite{gaoAcousticMetamaterialsNoise2022,moleronAcousticMetamaterialSubwavelength2015,kadic3d2019,padillaimaging2021,zhaoReviewPiezoelectricMetamaterials2022,kumaroverview2022,dongUnderwaterAcousticMetamaterials2022,zhangdiffusion2023}. 
This enables people to freely manipulate the propagation of waves. 
By combining transformation theory and the generalized Snell's law\cite{pendryControllingElectromagneticFields2006,leonhardtNotesConformalInvisibility2006,chenTransformationOpticsMetamaterials2010,yuLightPropagationPhase2011,sununderwater2023a}, wave modulation devices such as acoustic and optical cloaking\cite{liuBroadbandGroundPlaneCloak2009,zhuAcousticCloakingSuperlens2011,badrOpticalCarpetCloaking,chenBroadbandSolidCloak2017,biDesignDemonstrationUnderwater2017,sunQuasiisotropicUnderwaterAcoustic2019,xuPhysicalRealizationElastic2020}, metasurfaces\cite{maAcousticMetasurfaceHybrid2014,jiangConformalMetasurfacecoatedDielectric2017,arbabiMEMStunableDielectricMetasurface2018,renComplexamplitudeMetasurfacebasedOrbital2020,hePentamodeBasedCodingMetasurface2021}, and high-performance lenses\cite{lemoultResonantMetalensesBreaking2010,liDesignConformalLens2014,kainaNegativeRefractiveIndex2015,arbabiMEMStunableDielectricMetasurface2018,walkersubwavelength2020,padillaimaging2021} have been successfully realized. 
The successful experiences in optics and acoustics with metamaterials have provided insights for the development of controlling flexural waves in thin plates.

Flexural waves, one of the most prevalent elastic waves in thin plates, have garnered widespread and enduring attention in recent years. 
Researchers have achieved a diverse range of outcomes through the modulation of flexural waves, including innovations such as acoustic black holes, Luneburg lenses, and flexural wave metasurfaces. 
In as early as 1988, Mironov \cite{mironovpropagation1988} observed in wedge structures that when the thickness decreases in a power-law form ($h(x) = \varepsilon x^m, m\geq 2$), the wave velocity of flexural waves gradually decreases with decreasing thickness, ideally reaching zero velocity, achieving zero reflection of flexural waves. 
Krylov named such structures "acoustic black holes" (ABH) in 2002 \cite{krylovlaminated2002}, drawing parallels with their optical counterparts. 
This breakthrough not only paves the way for new research directions but also provides the opportunity for advancements in the theory and technology of structural vibration reduction and  energy harvesting\cite{zhaobroadband2014,zhaolowfrequency2019,dengreduction2020, shengvibration2023, wanmethod2023}.
Similar to acoustic black holes, multiple lenses have been designed to control flexural waves by varying the thickness of a thin plate. 
For instance, the design of Luneburg lenses achieve objectives such as flexural wave broadband beamforming\cite{zhaoflattened2020}, broadband ultralong subwavelength focusing\cite{zhaostructural2023}, cloaking and guiding\cite{zhaostructural2020}. 
Additionally, introducing structures on thin plates, typically by constructing a distributed array of slots or installing resonant pillars\cite{jingradient2019,caopillared2021,shenmetasurfaceguided2023c}, also can effectively modulate flexural waves. 
This approach finds widespread applications in flexural wave metasurfaces\cite{caodeflecting2018,jinelastic2021,caopillared2021,ruanreflective2021,wangexperimental2021}, topological modes\cite{wentopological2023} and flexural waveguides\cite{hurealization2022,shenmetasurfaceguided2023,shicompact2022}, among other fields.

The above-mentioned research enriches the functionality of flexural wave devices, broadens their application scenarios.
However, for a single flexural device, it also faces challenges such as limited functions, fixed application environments, and non-adjustable frequency bands. 
The emergence of programmable and reconfigurable metamaterial devices effectively addresses these issues\cite{meeussenGearedTopologicalMetamaterials2016,chenMultifunctionSwitchingFlat2020,chenMultifunctionSwitchingFlat2020,huengineering2023,weitransformable2023a,wangTunableUnderwaterLowfrequency2023}, enhancing the versatility of the devices and enabling their adaptation to diverse application scenarios.
Similarly, programmable and reconfigurable devices for flexural waves have also experienced rapid development.
Xu et al proposed an alternative method for beam splitting of flexural waves based on the generalized Snell's law, utilizing a coding meta-slab composed of two logical units, "0" and "1," which has been experimentally verified to effectively split incident flexural waves across a broadband frequency range\cite{xubeam2019}.
Yaw et al proposed a functionally switchable 3-bit active coding elastic metasurface, utilizing stacked piezoelectric patches and connected negative capacitance circuitry to achieve omnidirectional frequency control of elastic longitudinal waves, covering the entire $2\pi$ phase range and demonstrating a wide range of manipulation capabilities and high transmittance\cite{yawstiffness2021}.
Li et al introduced the concept of binary metasurfaces into the field of elastic waves, proposing a strategy for designing coding units through topology optimization. 
They successfully constructed a structured binary elastic metasurface, achieving accurate guidance and focusing of flexural waves, along with demonstrating an efficient energy harvesting system\cite{lisparse2022}.
Yuan et al. have designed and implemented a broadband reconfigurable metasurface using a "screw-and-nut" operating mechanism, achieving active control of the wavefront of flexural waves\cite{yuanreconfigurable2022}. 
Wu et al introduced a linear active metalayer, achieving independent frequency conversion of flexural waves in elastic beams and plates, opening up new possibilities for comprehensive control of time-domain signals and wave energy\cite{wuindependent2022}.
Recently, the Mikaelian lens has been utilized for achieving reconfigurable multifunctional achromatic focusing and imaging of flexural waves\cite{zhaobroadband2021,chenbroadband2022,chenultrabroadband2022a}. 
Zhao et al designed and experimental demonstrated a Mikaelian lens with large refraction to realize broadband sub-diffraction focusing without introducing evanescent waves\cite{zhaobroadband2021}.
Then a flexural wave Mikaelian lens was designed and fabricated by conformal transformation, successfully achieving achromatic subwavelength focusing (FWHM approximately $0.30\lambda$) and a beam-scanning angle of up to $120^\circ$ in the frequency ranges from $30$ kHz to  $180$ kHz. 
These work indicate significant potential applications of the Mikaelian lens in high-resolution medical imaging and flexural wave communication\cite{chenultrabroadband2022a}.
However, compared to its optical \cite{kotlyarSubwavelengthFocusingMikaelian2010,wangSelfFocusingTalbotEffect2017,chenconformally2021,zhounearfield2022} and acoustical counterparts\cite{SunZhaoYongJiYuGongXingBianHuanDeHuXingMikaelianShengTouJingSheJi2019,gaoConformallyMappedMultifunctional2019,sununderwater2023a}, research reports on flexural wave Mikaelian lenses are still relatively limited.
Therefore, the research on flexural wave Mikaelian lenses has important application value in designing and opening programmable and reconfigurable flexural wave devices based on Mikaelian lenses.

In this work, we design a flexural wave Mikaelian lens using conformal transformation, and use a piezoelectric array to implement active encoding communication based on the non-diffractive Talbot effect in the Mikaelian lens. 
The designed Mikaelian lens has a width of $W=280$ mm, a period of $l=880$ mm, and a length of $L=600\ \rm mm$, that is,  $68.18\%$ of the period. 
In order to reduce reflection, impedance matching layers with gradient thickness are designed at both ends of the lens. 
A piezoelectric sheet line array was arranged at the incident end of the lens as interference sources. 
The flexural wave Talbot effect generated by the interference sources are studied, and the non-diffraction characteristics of the Talbot effect in the Mikaelian lens are verified with simulation.
The interference source is imaged near the incident end of the lens.
Based on the self-focusing effect of the Mikaelian lens,  the image converges at one quarter of the period, and reappears as an inverted image near one half of the period. 
Therefore, if the channel state of the piezoelectric sheet is regarded as $1$ in the binary code, and the open one as $0$ in the binary code, then the incident signal can be regarded as being encrypted near the quarter period of the lens, and it is decoded again at the half of the period. 
Considering that the piezoelectric sheet interference source in this work is actively adjustable, active encoding of flexural waves can be achieved by our designed system. 
This research provides new ideas for the development of flexural wave programmable and reconfigurable multifunctional devices, and has potential value in flexural wave high-resolution imaging and communications.

\section{Theory and Design}
\begin{figure*}[hbt!]
    \centering
    \includegraphics[width=1\textwidth]{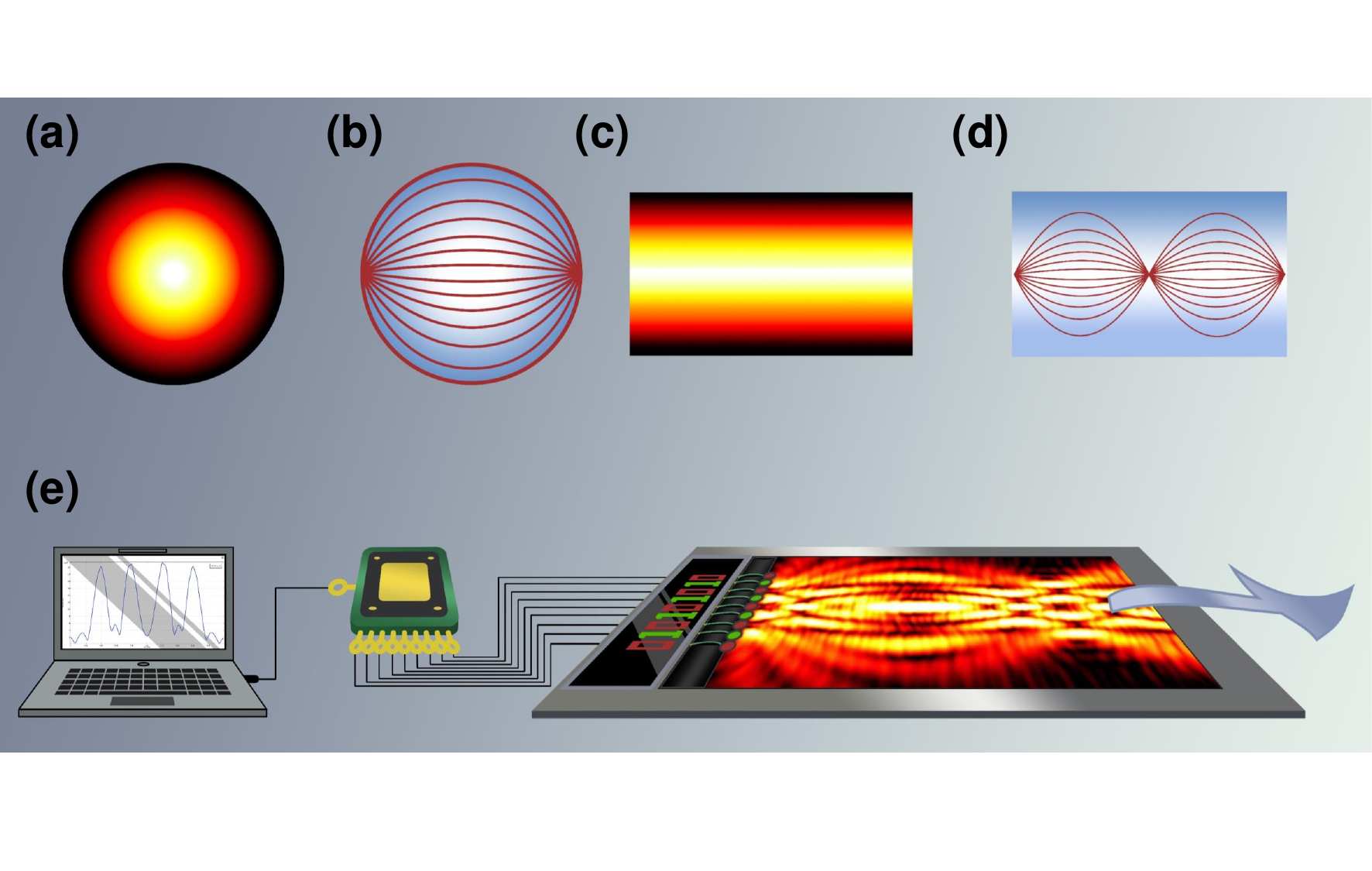}
    \caption{Theoretical design and communication encoding technique schematics of the conformal mapping Mikaelian lens . (a) Refractive index distribution of the Maxwell's fish-eye lens. (b) The rays trajectories in the Maxwell's fish-eye lens. (c) Refractive index distribution of conformal Mikaelian lens. (d) Self-focusing phenomenon of the Mikaelian lens. (e) Schematic representation of active encoding of flexural waves by the Mikaelian lens.}
    \label{fig:fig1}
\end{figure*}

The Mikaelian lens is a self-focusing lens designed by A.L. Mikaelian in 1951 \cite{mikaelianSelfFocusingMediaVariable1980}, which has a hyperbolic cosine inverse distribution of refractive indices. 
 Recent research indicates that the Maxwell's fish-eye lens and the Mikaelian lens can establish a mapping relationship through conformal transformations, as demonstrated below \cite{wangSelfFocusingTalbotEffect2017, gaoConformallyMappedMultifunctional2019}:
\begin{equation}
    n_z=n_w\left| \frac{dw}{dz} \right|,
    \label{eq:n_Z}
\end{equation}
where $n_z$ and $n_w$ are the refractive index distributions in virtual space (Maxwell's fish-eye) and  physical space (Mikaelian lens), respectively, as shown in Fig. \ref{fig:fig1}(a)-(d). 
The refractive index distribution for the Maxwell fisheye lens is:%
\begin{equation}
    n_w=\frac{2\alpha}{\left( 1+\left( \frac{r}{R} \right) ^2 \right)},
    \label{eq:n_w}
\end{equation}
where $\alpha$ and $R$ denote the refractive index on the lens boundary and the radius of the lens, respectively, and $r$ is the distance to the center of the lens, as shown in Fig.\ref{fig:fig1}(a).
The wave rays emanate from the left end of Maxwell's fish-eye lens and will then propagate along a circular arc and converge at the far right end of the lens, as shown in Fig.\ref{fig:fig1}(b).
Performing an exponential conformal transformation $w\left( u,v \right) =e^{\left( \beta z\left( x,y \right) \right)}$ on the Maxwell fish-eye lens yields a physical spatial refractive index distribution is:
\begin{equation}
  n(x,y) = n_0 \, \text{sech}\left(\frac{2\beta y}{W}\right)
\label{eq:n_xy}
\end{equation}
Equation (\ref{eq:n_xy}) represents the refractive index distribution of the Mikaelian lens. 
In Eq.(\ref{eq:n_xy}), $\beta$ is the length coefficient controlling the Mikaelian lens, $W = 2R = 280\ \rm {mm}$ represents the width of the lens , and $l = 2\pi R/\beta = 880\ \rm {mm}$ is the period of the Mikaelian lens.
And $n_0$ is the maximum refractive index along the lens's central axis,
defined as $n_0 = \alpha \beta$. %
The refractive index distribution of the lens, as shown in Fig. \ref{fig:fig1}(c), clearly indicates a gradual increase in refractive index from both sides of the lens towards its central axis. 
The self-focusing effect of the Mikaelian lens can be reflected through ray trajectories, as illustrated in Fig. \ref{fig:fig1}(d). 
In this work, we use Mikaelian lens to modulate the flexural wave and implement a kind of coding for flexural wave, where the schematic is shown in Fig. \ref{fig:fig1}(e).

\begin{figure*}[hbt!]
    \centering
    \includegraphics[width=1\textwidth]{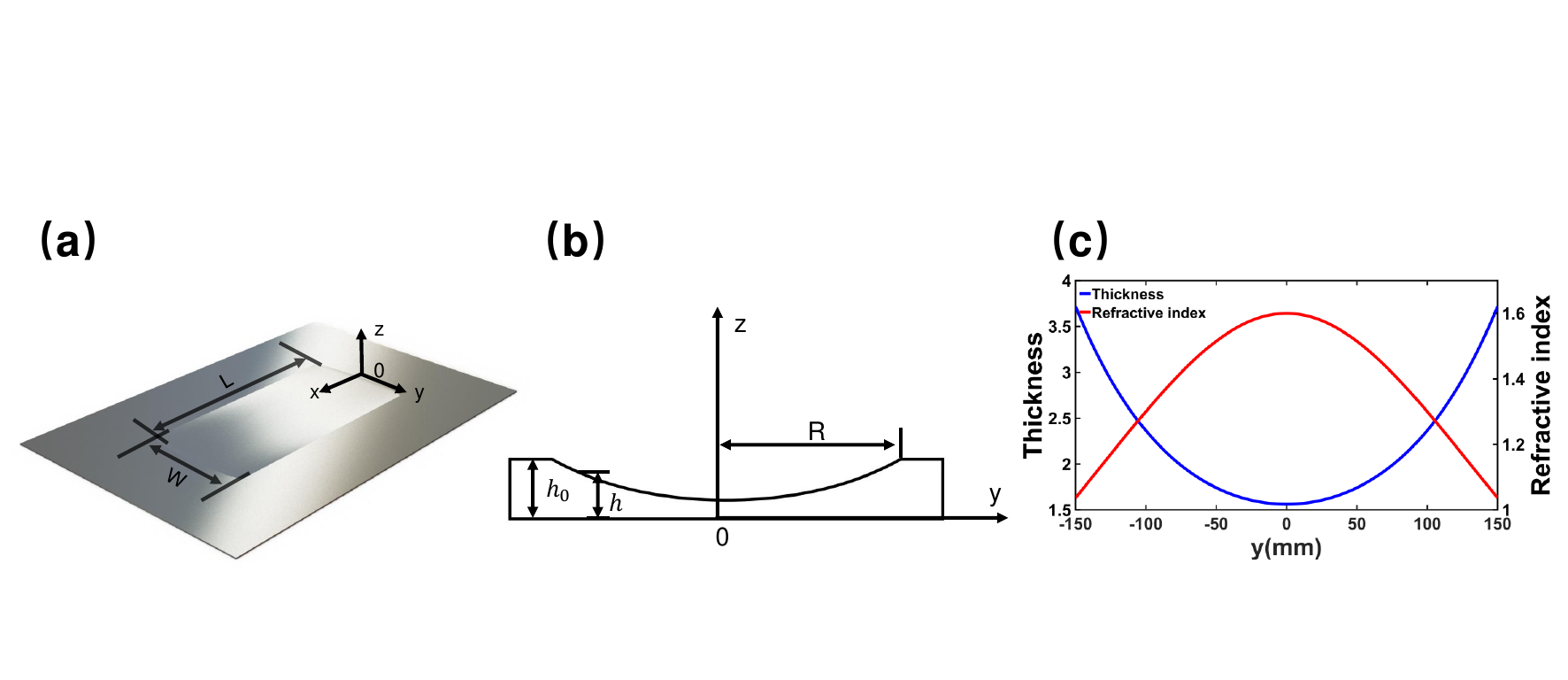}
    \caption{(a) The model of  the designed flexural wave Mikaelian lens. (b) Cross-sectional view of the lens with a thickness variation structure. (c) Relationship between thickness, refractive index, and distance from the lens center.}
    \label{fig:fig2}
\end{figure*}

The two-dimensional motion equation for flexural wave in a thin plate with variable local thickness $h\left( x,y \right)$ is:
\begin{equation}
\nabla ^2\left( D\nabla ^2w \right) -\left( 1-\upsilon \right) \cdot \left( \frac{\partial ^2D}{\partial y^2}\frac{\partial ^2w}{\partial x^2}-2\frac{\partial ^2D}{\partial x\partial y}\frac{\partial ^2w}{\partial x\partial y}+\frac{\partial ^2D}{\partial x^2}\frac{\partial ^2w}{\partial y^2} \right) +\omega ^2\rho h_w=0.
\label{eq:w_h}
\end{equation}
In Eq.(\ref{eq:w_h}),   $D= Eh^3\left( x,y \right) / 12\left( 1-\upsilon ^2 \right)$ indicates the flexural rigidity, $w$ represents the transverse displacement of the plate,$E,\rho, \upsilon, \omega$ and $h_w$ are  the Young's modulus, mass density, Poisson's ratio, angular frequency, longitudinal displacement of the plate correspondingly.
The solution of Eq.(\ref{eq:w_h}) is given as:
\begin{equation}
w = A\left( x,y \right) \,\,\exp \left( ik_v\phi \left( x,y \right) \right),
\label{eq:w_1}
\end{equation}
where $A\left( x,y \right)$ and $k_v \phi(x,y)$ are the wave amplitude and cumulative phase at the view point $\left( x,y \right)$.
Thus,  the velocity $c$ of flexural wave in a Mikaelian lens is expressed as:
\begin{equation}
c = \left( \frac{Eh^2\omega^2}{12\rho \left(1-\upsilon^2\right)} \right)^{\frac{1}{4}}.
\label{eq:c}
\end{equation}
The refractive index of the lens is controlled by the thickness of the localization \cite{chenbroadband2022}:
\begin{equation}
n(x, y) = \sqrt{\frac{h_0}{h(x, y)}}.
\label{eq:n_xy_2}
\end{equation}
As a result, the thickness distribution of the flexural wave Mikaelian lens can be expressed as:
\begin{equation}
h(x, y) = \frac{h_0\cosh^2\left(2\beta y/W\right)}{n_0^2}.
\label{eq:h_xy}
\end{equation}
The designed Mikaelian lens is embedded in a stainless steel plate with a thickness of $3.8\ \rm {mm}$, Young's modulus of $210\  \rm {GPa}$, density of $7800\ \rm {kg/m^3}$, and Poisson's ratio of $0.29$, as shown in Fig.\ref{fig:fig2}(a). 
To achieve impedance matching, an impedance matching gradient layer has been designed between the lens and the background plate. The cross-sectional view of the lens is illustrated in Fig.\ref{fig:fig2}(b), with the corresponding thickness and refractive index depicted in Fig.\ref{fig:fig2}(c).

\begin{figure*}[hbt!]
     \centering
     \includegraphics[width=0.85\textwidth]{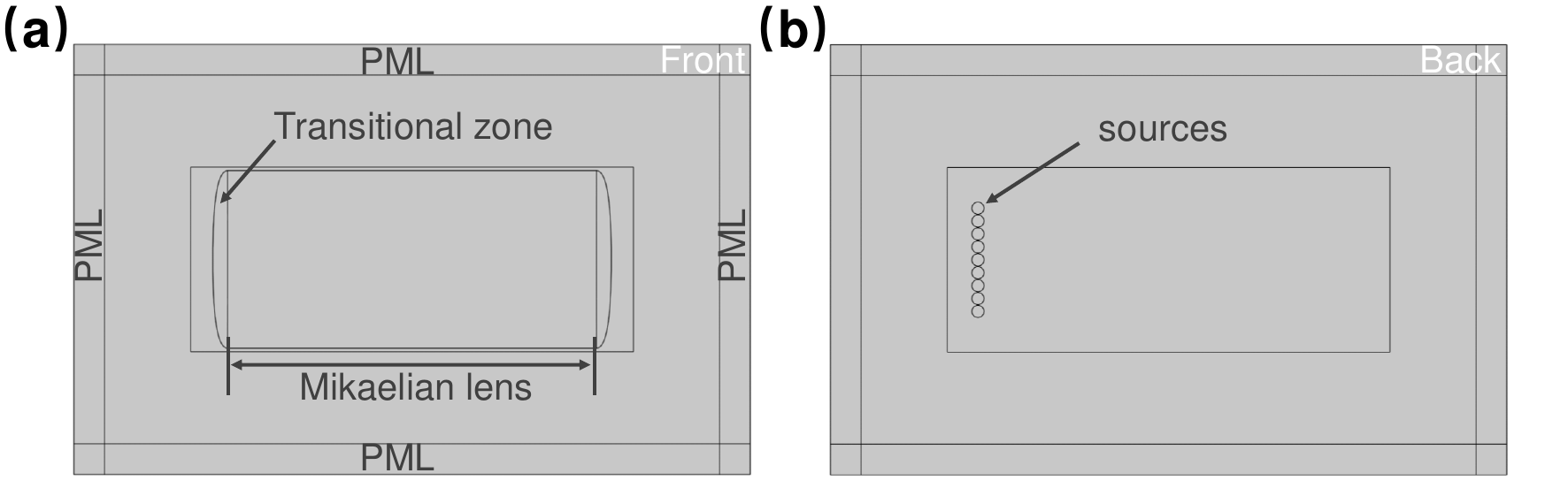}
     \caption{Schematic of the simulation model:(a) Schematic of the relative positioning of the Mikaelian lens on the stainless steel thin plate (front side).(b) Schematic of circular piezoelectric sheet array element (back side).}
     \label{fig:fig3}
\end{figure*}

The simulation model is shown in Fig.\ref{fig:fig3}(a), in which a transitional zone has been added between the lens and the thin plate\cite{zhaobroadband2021} to achieve impedance matching. 

The flexural wave is generated by an array of circular piezoelectric elements, shown as the circles in Fig.\ref{fig:fig3}(b), which can be expressed as:
\begin{equation}
w = \frac{A_0R_0}{r} \exp \left( ikr \right),
\label{eq:w_2}
\end{equation}
with $r_0$ is the radius of a circular source, $A_0$ representing the amplitude at a distance of $R_0$ from the source.
In addition, Perfect Mathc Layer (PML) are imposed to minimize the reflections.
Therefore, the superposition of $N$ sources on a point outside the plane can be explained as:
\begin{equation}
w = A_0r_0 \sum_{n=1}^N{\frac{\exp \left( i\left( kr_n+\phi_n \right) \right)}{r_n}},
\label{eq:w_3}
\end{equation}
where $\phi_n$ is the initial phase of each element, and $\varDelta \phi$ is the phase difference between adjacent elements, $r_n$ represents the distance from the $n$th element to the observation point. 
And the distance from the $n$th element to the first element is $\left( n-1 \right) d$ with $d$ as the distance between two adjacent sources. 
Therefore, $r_n$ in the equation \ref{eq:w_3} can be replaced by $r-\left( n-1 \right) d\sin \left( \theta \right) $, and $\phi_n$ can be replaced by $\left( n-1 \right) \varDelta \phi $, which can be rewritten as:
\begin{equation}
w=\frac{A_0r_0\exp \left( ikr \right)}{r}\exp \left[ -\frac{N-1}{2}\cdot i\cdot \left( kd\sin \left( \theta \right) -\varDelta \phi \right) \right] \cdot \frac{\sin \left[ \frac{N}{2}\left( kd\sin \left( \theta \right) -\varDelta \phi \right) \right]}{\sin \left[ \frac{1}{2}\left( kd\sin \left( \theta \right) -\varDelta \phi \right) \right]}.
\label{eq:w_4}
\end{equation}
By substituting $r=\sqrt{x^2+y^2}$, $\sin \left( \theta \right) =y/\sqrt{x^2+y^2}$ into the Eq.(\ref{eq:w_4}), it can be obtained that:
\begin{equation}
w=\frac{A_0r_0\exp \left( ik\sqrt{x^2+y^2} \right)}{\sqrt{x^2+y^2}}\exp \left[ -\frac{N-1}{2}\cdot i\cdot \left( \begin{array}{c}
	kd\frac{y}{\sqrt{x^2+y^2}}\\
\end{array} \right) \right] \cdot \frac{\sin \left[ \frac{Nkdy}{2\sqrt{x^2+y^2}} \right]}{\sin \left[ \frac{kdy}{2\sqrt{x^2+y^2}} \right]}.
\label{eq:w_5}
\end{equation}

\begin{figure*}[hbt!]
    \centering
    \includegraphics[width=1\textwidth]{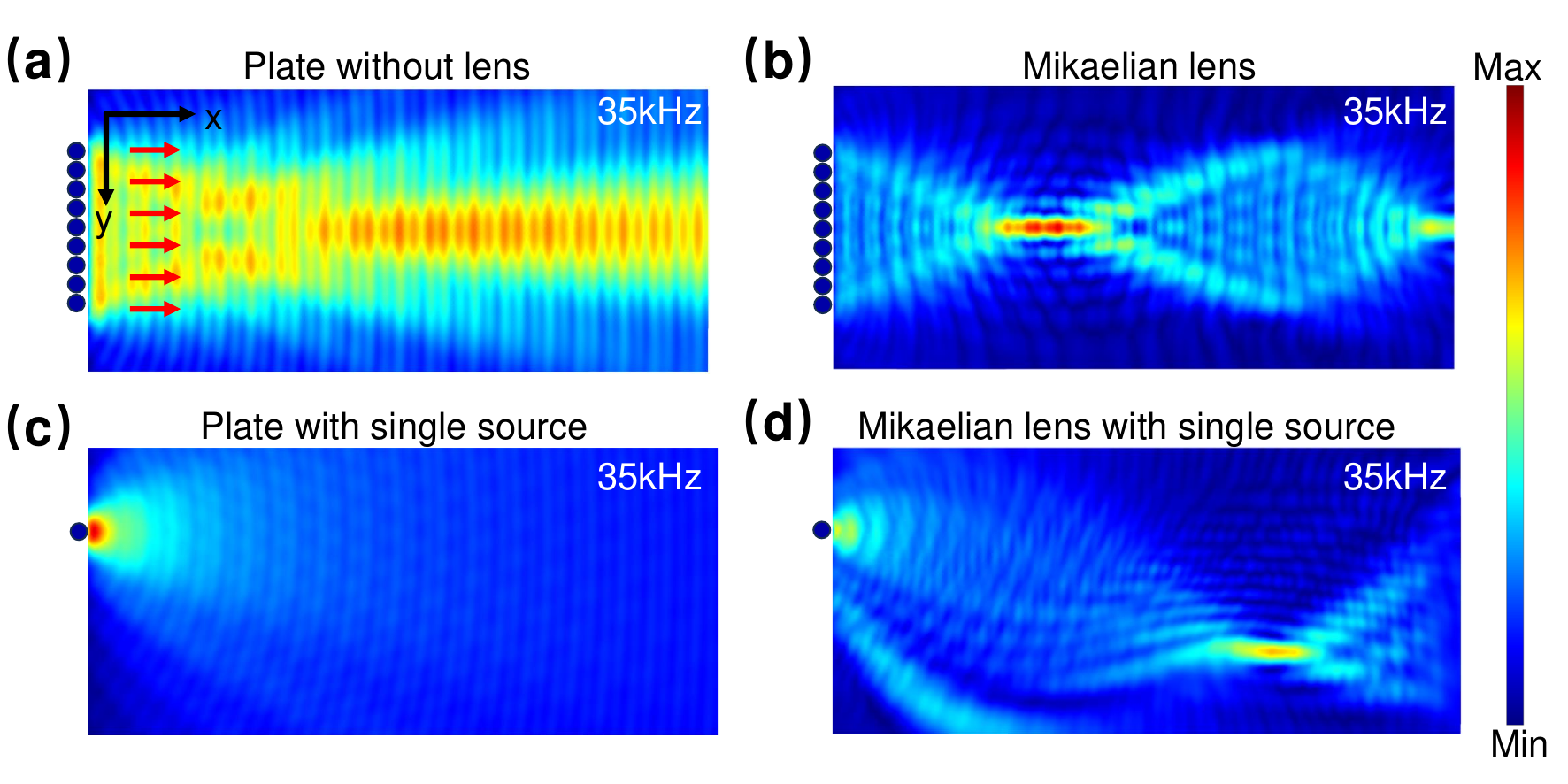}
    \caption{Simulation results of  displacement of plane wave in a thin plate (a) and in the designed  Mikaelian lens (b). Simulation results of the flexural wave excited by a single piezoelectric sheet propagating in a thin plate (c) and in the designed Mikaelian lens (d). }
    \label{fig:fig4}
\end{figure*}

According to Eq. (\ref{eq:w_5}), the modulation of the flexural wave propagation can be realized by modulating multiple piezoelectric sheet array element circles. 
If plane waves are to be generated in the far field, the initial phases of the piezoelectric elements in the array need to be the same, and the length of the array source can be controlled by the number of piezoelectric elements, as shown in Fig. \ref{fig:fig4}(a).

The equation for the trajectory of the rays inside a Mikaelian lens can be determined by the transverse components of the wave vector being equal to each other\cite{sununderwater2023a}:
\begin{equation}
   \frac{n\left( y\left( x \right) \right)}{\sqrt{1+y'^2\left( x \right)}}=n\left( y_0 \right) \cos \left( \theta \right).
    \label{eq:n_y(x)}
\end{equation}
Here, $y_0=y\left( 0 \right)$ represents the entry position on the $y$-axis to the left of the lens, and $\theta$ is the angle of incidence. 
The refractive index distribution of the Mikaelian lens mentioned above is given by Eq. (\ref{eq:n_xy}), thus the equation of the trajectory of the ray in the lens is given by\cite{suBroadbandFocusingUnderwater2017,lingradientindex2009,zigoneanuDesignMeasurementsBroadband2011}:
\begin{equation}
y\left( x \right) =\frac{W}{2\beta}\sinh ^{-1}\left[ \frac{\sinh \left( \frac{2\beta y_0}{W} \right) \cos \left( \frac{2\beta x}{W} \right)}{\cos \left( \theta \right)} \right].
\label{eq:y(x)}
\end{equation}
According to Eq. (\ref{eq:y(x)}), $y$ will be zero at $x=\pi W/4\beta$ in the case of plane wave normally incident to the lens. 
This means the rays converge at one-quarter of a period, as shown in Fig. \ref{fig:fig4}(b). 
Figure \ref{fig:fig4}(c) demonstrates the simulation results for the wave propagating in the ordinary thin plate, indicating that the flexural wave gradually spreads as it propagates. 
However, when the flexural wave excited by a single source is incident to the Mikaelian lens, as shown in Fig. \ref{fig:fig4}(d), there is a inverted image of this cousrce at $x=l/2$.
The point source appears as an inverted image at the $l/2$ position of the Mikaelian lens.

\newpage

\section{Non-diffractive Talbot Effect and Its Encoding in flexural Wave Mikaelian Lens}

\begin{figure*}[hbt!]
    \centering
    \includegraphics[width=0.75\textwidth]{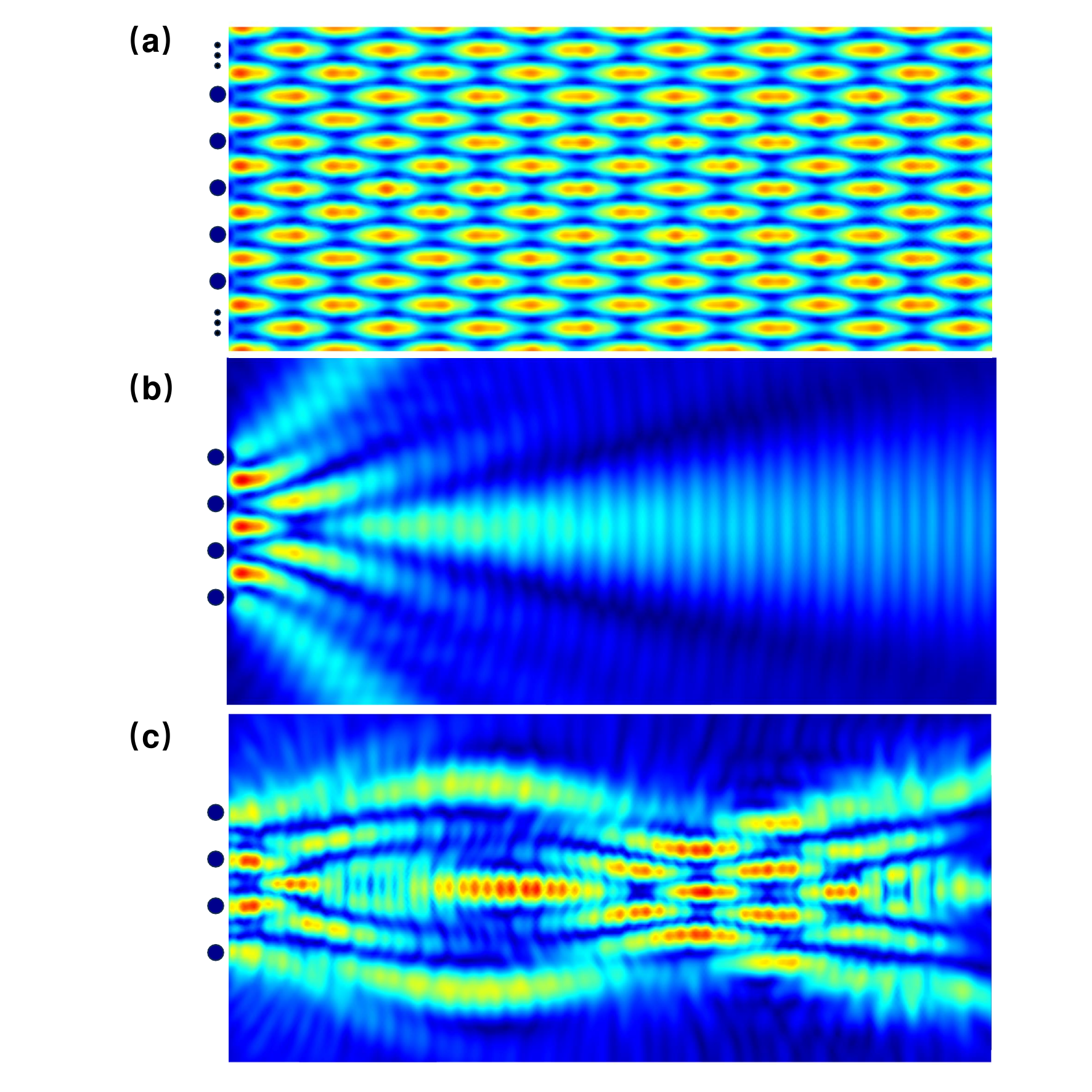}    
    \caption{(a) The Talbot effect generated by an infinitely periodic interference sources can propagate the image formed by the source to an infinite distance.(b) The Talbot effect generated by a finite periodic interference sources has a limited propagation distance.(c) The Talbot effect produced by a finite periodic interference sources exhibits non-diffractive propagation behavior within the Mikaelian lens.} 
    \label{fig:fig5}
\end{figure*}

The Talbot effect was first found by Talbot in optics\cite{talbotFactsRelatingOptical1836}, and has practical applications in optical fields for beam collimation tests and precise lens focal length measurements\cite{wenTalbotEffectRecent2013}. 
Considering the similarity between optical wave and elastic wave, the Talbot effect should also appear in flexural wave.
Since the phenomena of diffraction and interference are all produced by the wave's superposition, the periodic interference sources can also generate the Talbot effects.
Figure \ref{fig:fig5}(a) shows the interference field produced by the infinite periodic sources with a period of $D=4r_0$.
It can be observed that periodic sources repeat along the propagation direction at integer multiples of the main Talbot length $2D^2/\lambda$ and are equally spaced in the transverse direction. 
This has the same phenomena of Talbot effect produced by infinite periodic diffraction gratings.
However, in practical systems, the array cannot be infinitely large, Fig.\ref{fig:fig5}(b) displays the simulated results of the ordinary Talbot effect for finite-period sources. 
Clearly, due to the diffraction, the Talbot effect can only be maintained over relatively short distances in finite-period systems. 
Consequently, it is not possible to achieve long-distance propagation of the sources' modes through the Talbot effect in thin plate. 
Since the Mikaelian lens has a self-focusing effect, it can achieve a non-diffractive Talbot effect. 
Theoritically, the non-diffractive Talbot effect of the Mikaelian lens can propagate the modes of the interference sources over a considerable distance. 
Figure.\ref{fig:fig5}(c) shows the simulation results of the Talbot effect propagating in the Mikaelian lens, which is excited by four adjacent piezoelectric plates with a spacing of $4 \ \rm cm$.
It can be observed that the Talbot diverge at a quarter of the period but reimage at half the period. 
This mode will periodically repeat as the propagation distance increases.
\begin{figure*}[hbt!]
     \centering
     \includegraphics[width=0.95\textwidth]{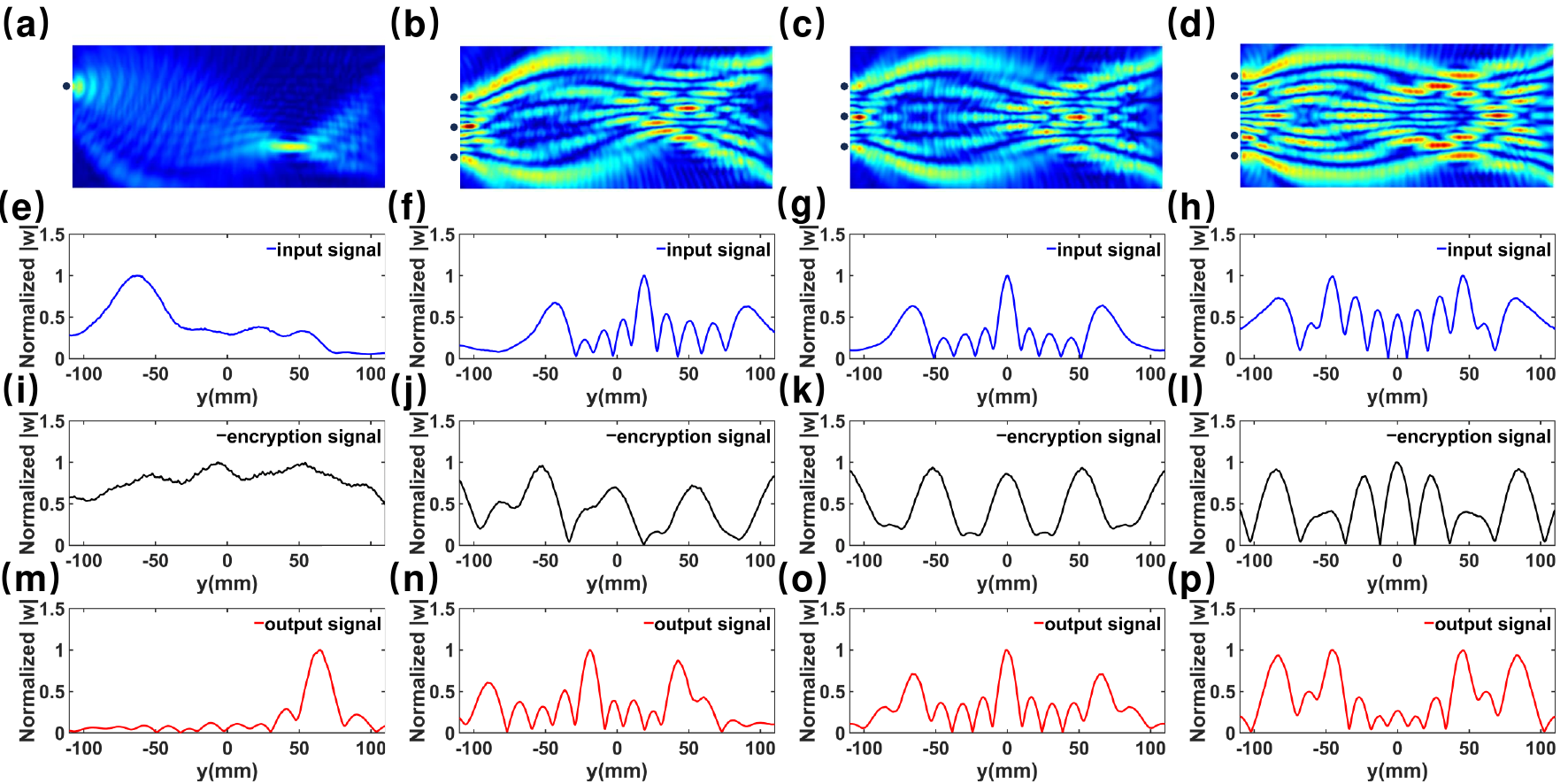}
     \caption{Simulation results of the displacement field for a single interference source (a),  sources of $001001001$ (b), sources of $010010010$ (c) and sources of $101000101$ (d).
     Normalized  displacement distribution of the input signals (e)-(h), encryption signals (i)-(l) and output signals (m)-(p), corresponding to (a)-(d), respectively.}
     \label{fig:fig6}  
\end{figure*}

The research above demonstrates that the Talbot effect produced by finite-period interference sources in the Mikaelian lens possesses non-diffractive properties and can reform the image of the interference source at the half-period position. 
This feature can be used to achieve communication encoding for flexural waves. 
To further investigate the potential of using the non-diffractive Talbot effect of flexural waves in the Mikaelian lens for communication encoding, simulations are conducted using the commercial software COMSOL Multiphysics. 
The simulations are performed at $35$ kHz with the specified displacements applied at the incident end of the Mikaelian lens model (Fig.\ref{fig:fig3}(b) sources position), and the  displacement field was observed both on the thin plate and the back of the lens. 
By adjusting whether the actively controlled piezoelectric elements were excited, encoding was performed using "0" for no excitation and "1" for excitation. 
The simulation results are shown in Fig.\ref{fig:fig6}, where Fig.\ref{fig:fig6}(a), (b), (c), (d) represent the displacement fields of flexural waves for various encoding sequences of $01000000$, $001001001$, $010010010$, and $101000101$, respectively. 
The input signal (blue solid line) is obtained near the left end of the Mikaelian lens.
The encrypted result (black solid line) is at the $x = l/4$ (focusing plane).
The final output result (red solid line) is obtained at $x = l/2$. 
In this paper, 9-bit encoding is used to investigate the ability of the Talbot effect to encode information in the conformation Mikaelian lens, and the encoding information was transmitted and encrypted in the lens.
It can be observed that the four sequences are transmitted and focused over long distances in the lens. 
Figure.\ref{fig:fig6}(e) shows the  normalized displacement distribution of the input signal when exciting a single piezoelectric element; Fig.\ref{fig:fig6}(i) shows the  normalized displacement  distribution of the encrypted result; Fig.\ref{fig:fig6}(m) shows the  normalized displacement  distribution of the output signal. Figure.\ref{fig:fig6}(f)(j)(n), (g)(k)(o), (h)(l)(p) correspond to different encoding sequences.
The simulation results show that the designed system has a good performance for non-diffractive Talbot effect encoding.

\section{Experimental Results}
\begin{figure*}[hbt!]
    \centering
     \includegraphics[width=0.9\textwidth]{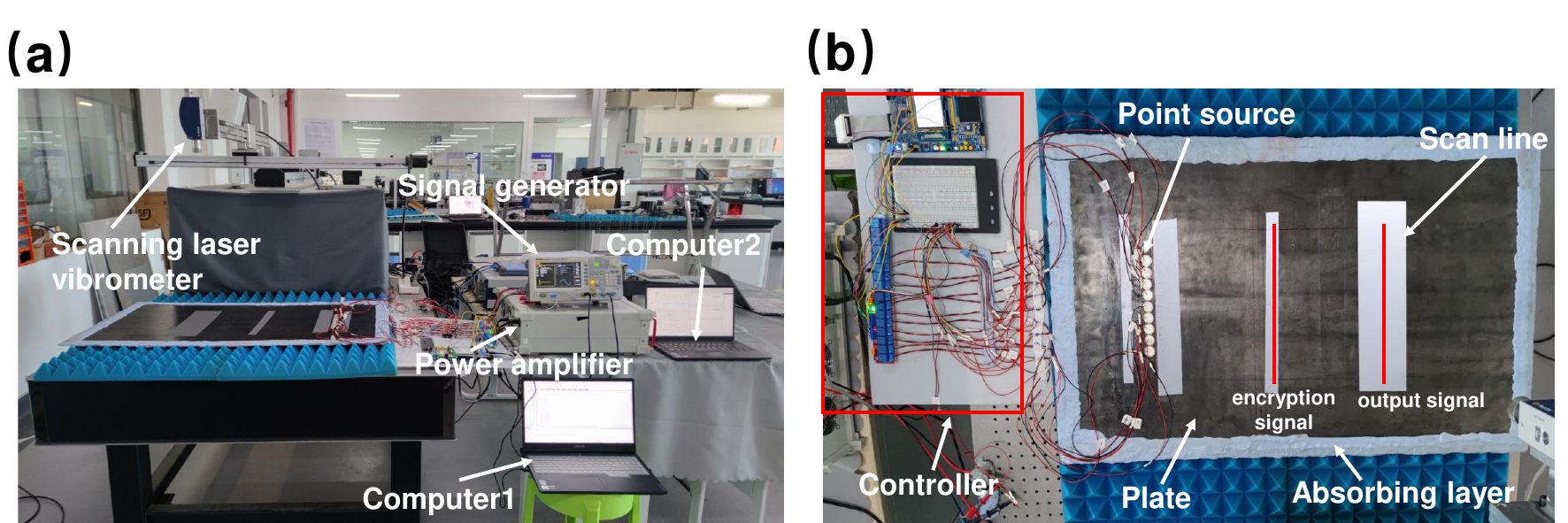}
    \caption{(a) Experimental setup.(b) The wave source is mounted on the backside of the sample, and the displacement normalized distribution of the lens's focusing and imaging positions is measured using a laser vibrometer. 
    }
    \label{fig:fig7}
\end{figure*}

Based on the simulation results presented above, Mikaelian lens sample was machined from a thin stainless steel plate using laser selective fusion to validate the capability of using the self-focusing phenomenon of flexural wave self-focusing and non-diffractive Talbot communication.
The experimental setup, as shown in Fig.\ref{fig:fig7}, consists of the flexural wave Mikaelian lens, a signal generator, an array of piezoelectric elements with a radius of $r_0=10\ \rm {mm}$, a controller comprising a Stm32 microcontroller and relays, a laser scanner, and other components. 
The Mikaelian lens is surrounded by blu-tack to absorb reflected waves. 
Nine circular piezoelectric elements are attached to the back of the lens and are located at the incident end of the lens to generate incident flexural waves. 
The incident wave signal applied to the piezoelectric elements is generated by a signal generator and amplified by a power amplifier. 
The  displacements are scanned and recorded by a laser vibrometer.
\begin{figure*}[hbt!]
   \centering
     \includegraphics[width=1\textwidth]{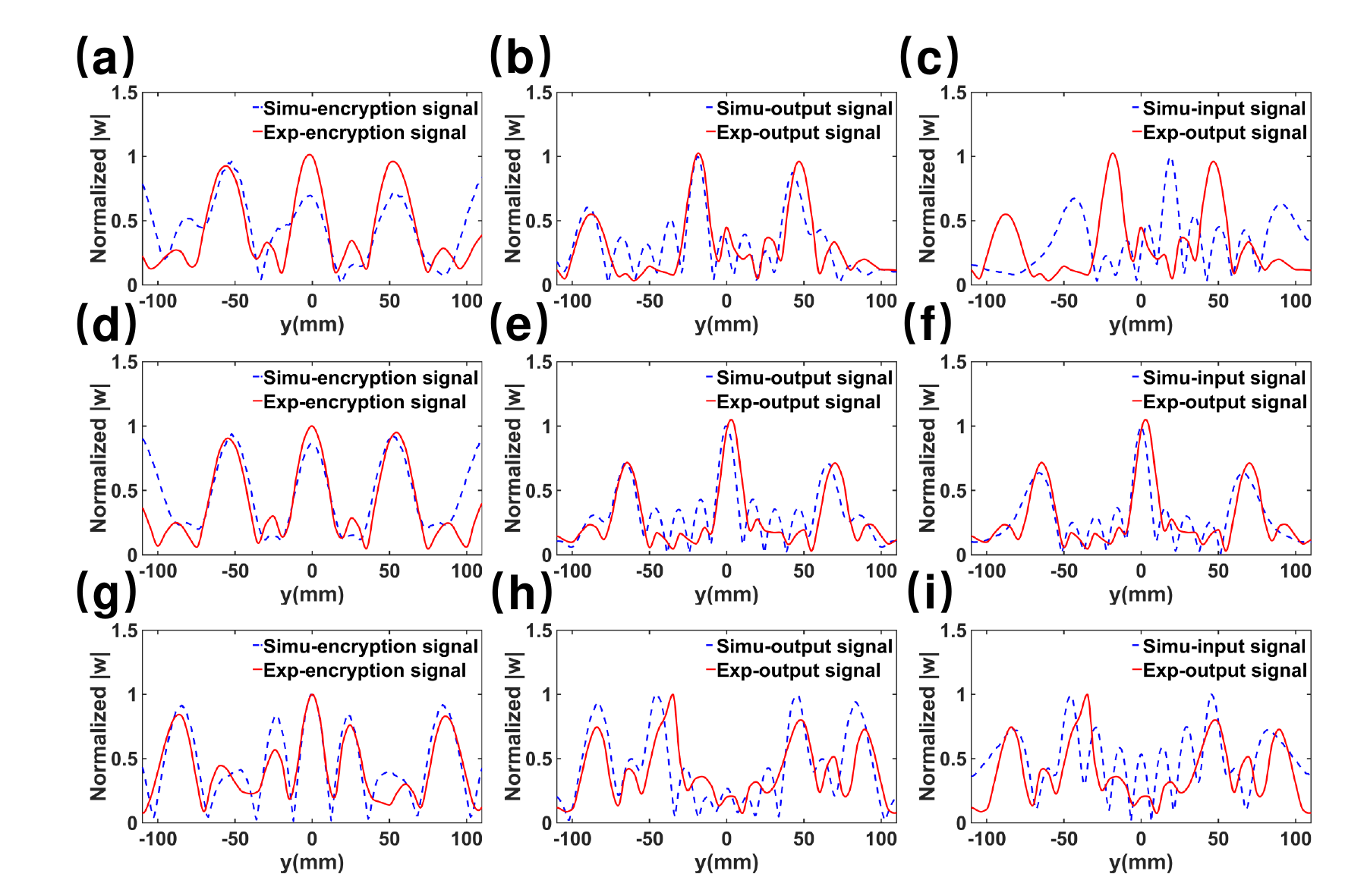}
     \caption{ The simulations (dashed blue curves) and experiments (solid red curves) of the non-diffractive encodings of $001001001$ (a)-(c), $010010010$ (d)-(f) and $101000101$ (g)-(i), respectively.
     The normalized   displacement distributions of the encryption signals ((a),(d) and (g)) and output signals ((b),(e) and (h)).
     Comparison of the input signals and output signals of the three coding sequences of $001001001$ (c), $010010010$ (f) and $101000101$ (i), respectively.} 
     \label{fig:fig8}
\end{figure*}

The controller has multiple channels to independently control their different element of the line array.
The open and closed circuit states represent the binary codes of $0$ and $1$, respectively
A signal generator is used to excite the circular piezoelectric elements. 

Figure \ref{fig:fig8} shows the simulated and experimental results of the non-diffractive encoding, where three different $9\-$bits codes, i.e., $001001001$, $010010010$ and $101000101$, are  employed to verify the encoding capacities.
Figure \ref{fig:fig8}(a)-(c) are the cases of $001001001$, where in Fig.\ref{fig:fig8}(a) the dashed and solid lines represent the displacement distribution of the simulation and experiment at $x=l/4$, respectively.
It can be seen that the simulated results match with the experiments well.
The output signals are shown in Fig.\ref{fig:fig8}(b), where the simulated and experiment result have a high consistency.
Since the input and output signals are near the left end and half of the period of the Mikaelian lens correspondingly, they are anti-symmetric about $y=0$.
This can be further proved in Fig.\ref{fig:fig8}(c), where the dashed and solid curves are the input and the output signals, respectively.
According to Fig.\ref{fig:fig8}(a)-(c), it is obvious  that the system works well for the encoding of $001001001$.
Figure \ref{fig:fig8}(d)-(f) and (h)-(i) are the experiment results of the cases of $010010010$ and $101000101$, respectively.

\begin{figure*}[hbt!]
    \centering
     \includegraphics[width=0.9\textwidth]{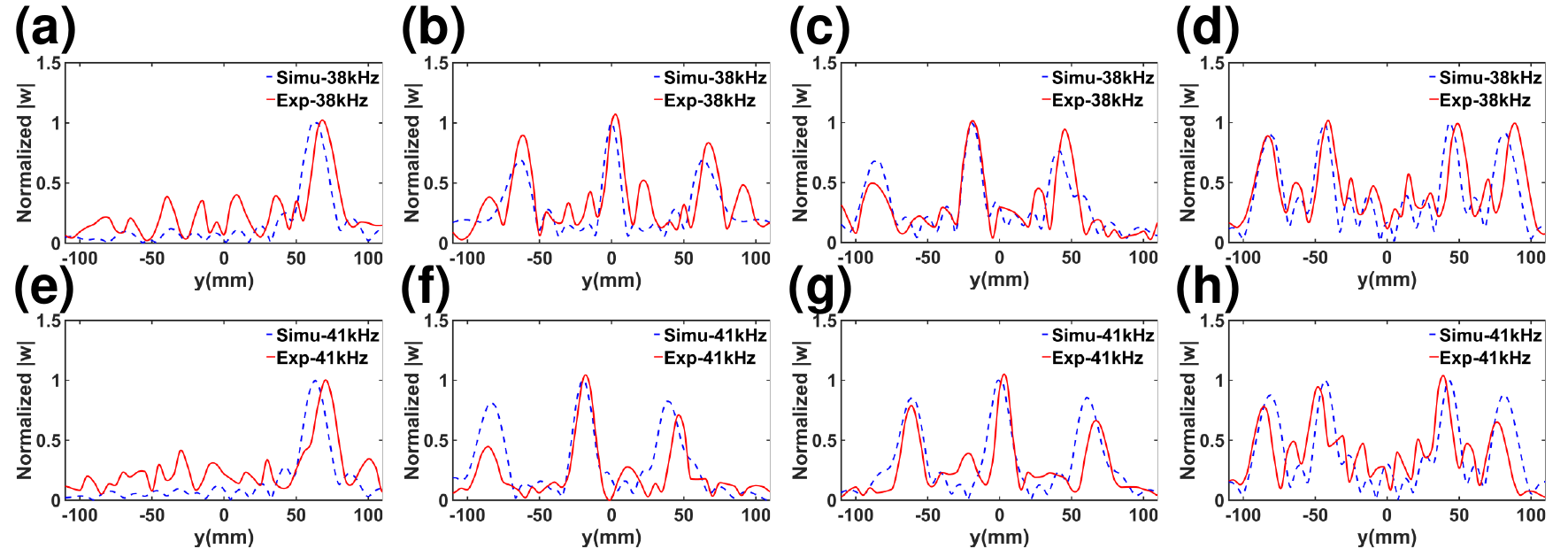}
    \caption{(a) and (e) are the comparison charts of the experimental (red solid line) and simulated (blue dashed line) normalized  displacement distributions along the y-direction at the incidence and imaging positions when a single piezoelectric element is excited at 38 kHz and 41 kHz.(b)(f), (c)(g), and (d)(h) correspond to the three encoding sequences: 001001001, 010010010, and 101000101 at $38\ \rm kHz$ and $41\ \rm kHz$, respectively.} 
    \label{fig:fig9}
\end{figure*}

Since the encoding process is based on the active interference sources, the system works in a broadband frequency range.
This is further proved by the experiments, as shown in Fig.\ref{fig:fig9}, where the simulated and experimental displacements at output ends  are represented by the dashed and solid curves correspondingly.
Figure \ref{fig:fig9}(a) shows the displacements at the output ends produced by a single source at $38\ \rm kHz$, where there can be found the good matching between the simulations and experiment.
And Fig.\ref{fig:fig9}(b) to (d) are the cases of the coding sequences of $001001001$,$010010010$ and $101000101$ at $38\ \rm kHz$ respectively, in which the simulations matches the experiments well.
Figure \ref{fig:fig9}(e)-(f) are the $41\  \rm kHz$ counterparts of the cases of Fig.\ref{fig:fig9}(a)-(d).

The encoding process relies on active interference sources, enabling the system to operate across a broad frequency spectrum. 
This is corroborated by experimental results illustrated in Fig. \ref{fig:fig9}, where dashed and solid curves represent simulated and experimental displacements at the output ends, respectively. 
In Fig. \ref{fig:fig9}(a), displacements at the output ends, generated by a single source at $38\ \rm kHz$, exhibit excellent agreement between simulations and experiments. 
Figures \ref{fig:fig9}(b) to (d) correspond to coding sequences of $001001001$, $010010010$, and $101000101$ respectively, demonstrating consistent alignment between simulations and experiments. 
Figures \ref{fig:fig9}(e)-(f) present the $41\ \rm kHz$ counterparts of the scenarios in Fig. \ref{fig:fig9}(a)-(d).
Additionally, it is noteworthy that the performance remains robust at $41\ \rm kHz$, as indicated by the close agreement between simulations and experiments in Figures \ref{fig:fig9}(e)-(f).

\begin{figure*}[hbt!]
    \centering
     \includegraphics[width=0.9\textwidth]{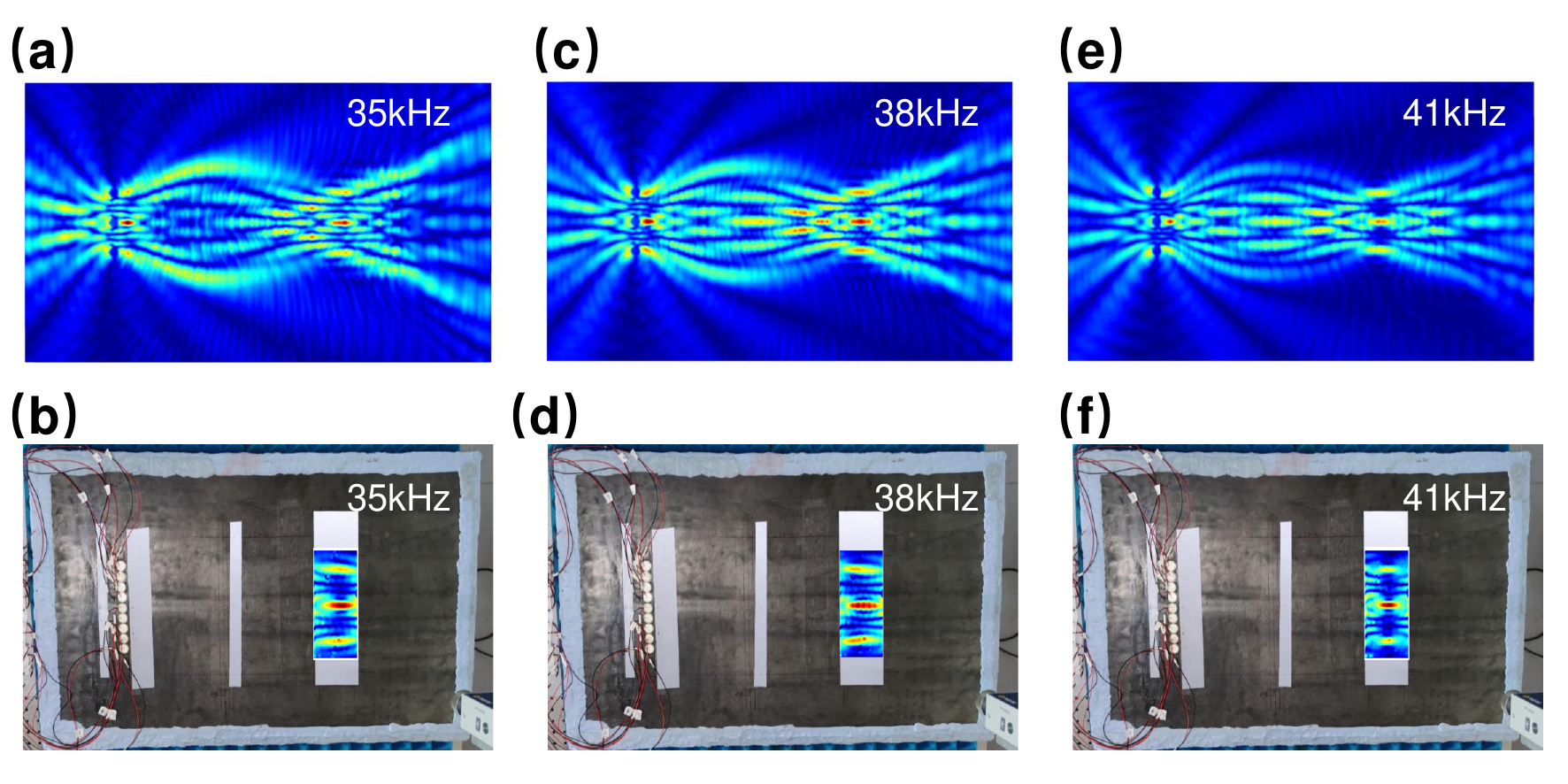}
    \caption{The simulated displacements of the coding sequences of $010010010$ at (a) $35\ \rm kHz$, (b) $38\ \rm kHz$ and (c) $41 \ \rm kHz$. The experimental displacements of the coding sequences of $010010010$ at(d) $35\ \rm kHz$, (e) $38\ \rm kHz$ and (f) $41 \ \rm kHz$.}
    \label{fig:fig10}
\end{figure*}

In addition, we conduct a field scan of the output displacements of the system at 35 kHz, 38 kHz, and 41 kHz to visually illustrate the alignment between simulations and experiments, as depicted in Fig.\ref{fig:fig10}. 
Specifically, Fig.\ref{fig:fig10}(a)-(c) represent the simulated displacement fields for the coding sequence $010010010$ at $35 \ \rm kHz$, $38\ \rm kHz$, and $41 \ \rm kHz$, respectively. 
It is evident that the Talbot effect can be generated through the interference sources, and a reconstructed image of the source is formed at $l/2$ after convergence at the coding output end. 
Corresponding experimental results are shown in Fig.\ref{fig:fig10} (d)-(f), where the displacement field was measured in an area near the output end with a width of $80$ mm and a length of $200$ mm, with a scanning step size of $5$ mm. 

The high consistency between experimental and simulated results confirms the broad-frequency effectiveness of the designed system.

\section{Conclusions}
This article presents the design of a Mikaelian lens for manipulating flexural waves, featuring a lens period of $880\ \rm mm$, a width of $280\ \rm mm$, and constructed from stainless steel. 
Utilizing finite element simulation, the study investigates the propagation characteristics of flexural waves within the lens. 
Simulation results demonstrate consistency with the wave trajectory equation. 
When piezoelectric patches with a radius of $10 \ \rm mm$ serve as interference sources, an array of periodically arranged interference sources generates the Talbot effect for flexural waves. 
Unlike the Talbot effect produced by finite-period interference sources in ordinary thin plates, which is confined to the near field due to diffraction, the research indicates that, for the Mikaelian lens, the self-focusing function allows the Talbot effect to propagate to more distant locations, exhibiting non-diffractive characteristics. 
Additionally, controlling the on-off states of interference sources can manipulate the distribution of bright spots in the Talbot effect. 
Exploiting this feature, an active encoding method is proposed. 
In this method, the quarter-period of the lens serves as the encryption end, and leveraging the self-focusing effect of the lens, reimaging occurs at the half-period, achieving active encoding of flexural waves. 
Simulation results, validated by experimental findings, affirm the effectiveness of active encoding. 
This study holds potential applications in flexural wave communication, detection, and other domains.

\section*{Acknowledge}
The authors sincerely acknowledge the financial support of BIGC Projects (Grant Nos. BIGC Ed202206, 27170123007), the Project of Construction and Support for high-level Innovative Teams of Beijing Municipal Institutions(Grant No. BPHR20220107), the financial support of Anhui Provincial Natural Science Foundation (Grant No. JZ2023AKZR0583).


\end{document}